\documentclass{article}
\usepackage{spconf,amsmath,graphicx,hyperref}
\usepackage{amssymb}
\usepackage{booktabs}
\usepackage{enumitem}
\usepackage{svg}
\usepackage{cite}
\title{SE-ADD: Self-Evolving Audio Deepfake Detection with Mistake-Driven Supervision}
\name{Rong Wan$^{1}$, Wei Xie$^{2}$, Jiaxi Li$^{1}$, Wenwu Wang$^{1}$, Lu Yin$^{3}$, Yiliao Song$^{4}$, Xilu Wang$^{1}$\thanks{This work has been submitted to the IEEE for possible publication.
Copyright may be transferred without notice, after which this version may no longer be accessible.}
}
\address{$^{1}$ University of Surrey, $^{2}$ Guangxi University,\\$^{3}$ Shenzhen University of Advanced Technology, $^{4}$ Adelaide University}

\begin{document}

\maketitle
\begin{abstract}
Audio deepfake detection (ADD) must remain effective when new spoofing attacks emerge after deployment. 
Emerging audio language model (ALM)-based ADD methods are built on predefined supervision from ground-truth labels or verified forensic rationales.
However, this paradigm overlooks an ALM's own mistakes, which indicate where targeted supervision is most needed. 
To this end, we first introduce evolving spoofing environments for ALM-based ADD, where a new attack becomes dominant while previously observed attacks persist. 
Motivated by the above learning-from-mistakes perspective, we further propose SE-ADD, a self-evolving framework that iteratively adapts an ALM via low-rank adaptation (LoRA) using mistake-driven supervision built from its verdicts and self-generated forensic cues. 
All training samples receive direct authenticity supervision, while misclassified ones receive additional cue-augmented supervision. 
As verdicts and cues are regenerated by the updated ALM, the resulting supervision evolves accordingly.
Experiments on two ALMs demonstrate the effectiveness of SE-ADD in generalizing to unseen attacks, reducing the equal error rate (EER) from $36.72\%$ to $7.52\%$ for Qwen2-Audio and from $19.93\%$ to $3.97\%$ for MOSS-Audio.

\end{abstract}
\section{Introduction}
\label{sec:intro}

Evolving speech-generation technologies continually introduce new dominant spoofing attacks~\cite{SDD-survey}. This poses a challenge for audio deepfake detection (ADD), which must remain effective as spoofing attacks change~\cite{ML-ITW}. 
Due to the promising performance of audio language models (ALMs) across various audio reasoning tasks~\cite{Audio-CoT,AIR-Bench}, recent studies have begun to explore ALM-based ADD methods. These approaches primarily rely on predefined supervision, such as ground-truth labels~\cite{allm4add} or curated forensic reasoning in the form of rationales~\cite{HIR-SDD} and chain-of-thought annotations~\cite{CoLMbo-DF}.

However, such a paradigm is specified independently of an ALM's current failures in ADD, overlooking its potential to learn from its own mistakes. These failures can reveal limitations in the model's current ADD capability and indicate where additional supervision would be beneficial~\cite{an2024learningmistakesmakesllm,tong-etal-2024-llms}. Moreover, such supervision remains fixed despite changes in an ALM's performance as the spoofing environment evolves. This leaves open whether an ALM can use its own mistakes to guide additional supervision as the dominant spoofing attack changes. To study this problem, we introduce a practically motivated evolving ADD setting composed of a sequence of spoofing environments. In each environment, the training and development sets share the same evolving attack mixture and contain balanced bona fide and spoof samples, with a newly introduced attack dominating the spoof samples while previously observed attacks remain at low prevalence. The development set is used to select the best-performing adaptation within the current environment. Separately, an evaluation set, which consists of bona fide samples and spoof samples generated by unseen attacks, is used to assess generalization.

Meanwhile, self-evolving language models suggest that model-generated outputs can serve as training signals for subsequent parameter updates~\cite{seal,lance,pmlr-v235-chen24j,pmlr-v235-yuan24d}.
Combining the self-evolving mechanism with the learning-from-mistakes perspective, we propose self-evolving audio deepfake detection (SE-ADD), a framework for ALM-based ADD that performs iterative low-rank adaptation (LoRA)~\cite{hu2021lora} within each evolving environment using mistake-driven supervision.
At each round, the current ALM generates forensic cues and cue-free predictions on the training set. All training samples receive direct authenticity supervision, while misclassified ones additionally receive cue-augmented supervision using the forensic cues generated by the ALM. 
Within each environment, the updated ALM regenerates cues and predictions for the next round. Selected on the development set, the best-performing LoRA adaptor is carried forward to the next environment.
\begin{figure*}[!t]
    \centering
    \includegraphics[width=\textwidth]{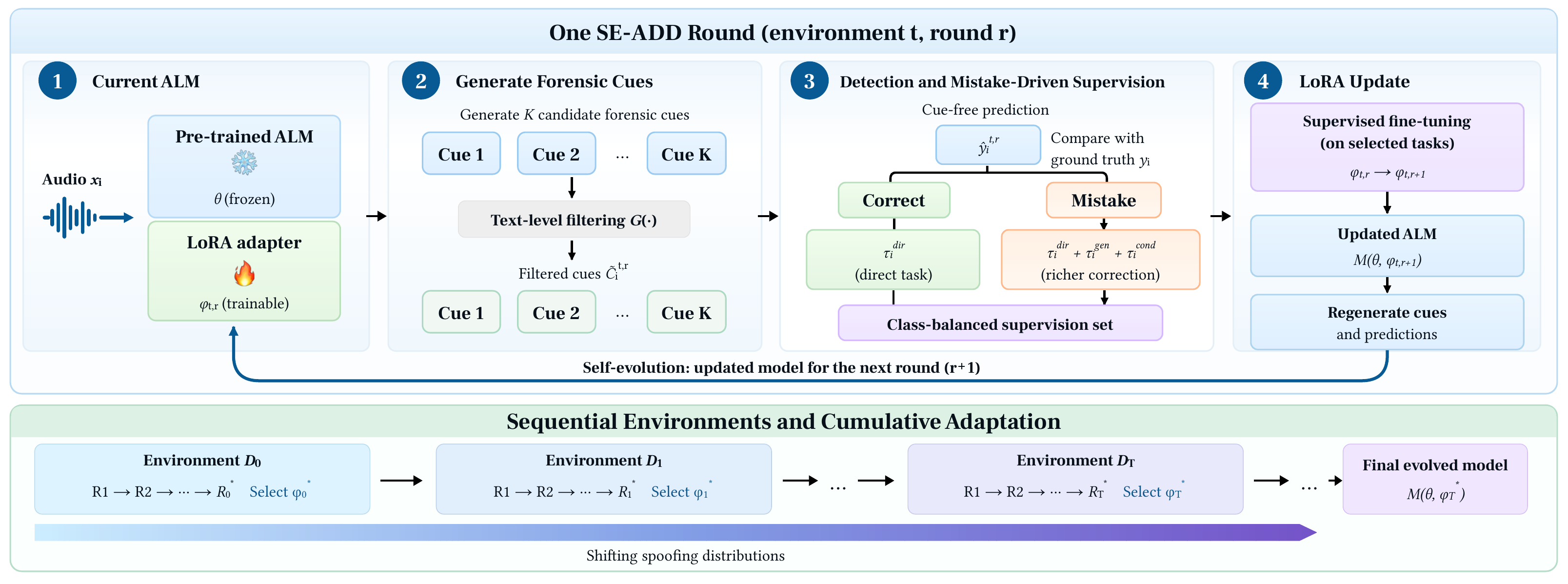}
    \caption{Overview of SE-ADD.
Top: one self-evolution round with forensic-cue generation, mistake-driven supervision, and LoRA updating.
Bottom: selected LoRA parameters are carried across shifting environments
for cumulative adaptation.}
    \label{fig:seadd}
\end{figure*}

In summary, our contributions are threefold:
\begin{itemize}[leftmargin=1.2em,labelsep=0.5em,itemsep=1pt,topsep=2pt,parsep=0pt]
    \item We introduce ALM-based ADD under evolving spoofing environments, where new attacks become dominant while previously observed attacks remain present.

    \item We propose SE-ADD\footnote{\url{https://github.com/SE-ADD/SE-ADD}}, which uses ALM-generated forensic cues and its detection failures to construct mistake-driven supervision for cumulative LoRA adaptation.

    \item We evaluate SE-ADD on two ALMs across six evolving environments, showing consistent adaptation, transfer to held-out unseen attacks, and effective forensic-cue supervision for cue-conditioned detection.
    
\end{itemize}

\section{Problem Setup}
\label{sec:problem}

ALM-based ADD is considered under a sequence of $T$ evolving spoofing environments $\{\mathcal{E}_t\}_{t=0}^{T-1}$, where the relative prevalence of spoofing attacks changes over time.
Each environment provides a labeled training set $\mathcal{D}^{\mathrm{train}}_t=\{(x_i,y_i)\}_{i=1}^{N_t}$, where $x_i$ is an audio utterance and $y_i\in\{\textit{bonafide},\textit{spoof}\}$.
A disjoint development set $\mathcal{D}^{\mathrm{dev}}_t$ is used for model selection within the same environment.

We denote the ALM at evolution round $r$ of environment $t$ by
$\mathcal{M}(\theta,\phi_{t,r})$, where $\theta$ contains the frozen
base-model parameters and $\phi_{t,r}$ the trainable LoRA parameters.
Here, $r=0$ denotes the model entering an environment, and the
within-environment evolution carried only by $\phi_{t,r}$ is
\begin{equation}
    \mathcal{M}(\theta,\phi_{t,0})
    \rightarrow
    \mathcal{M}(\theta,\phi_{t,1})
    \rightarrow
    \cdots
    \rightarrow
    \mathcal{M}(\theta,\phi_{t,R}),
    \label{eq:model_evolution}
\end{equation}
where $R$ is the maximum number of evolution rounds.

To model the distribution shift, let $\mathcal{A}=\{a_i\}_{i=1}^{A}$ denote the set of spoofing attacks and define the spoof-attack mixture in environment $\mathcal{E}_t$ as $\pi_t(a_i)=P(A=a_i \mid Y=\textit{spoof},\mathcal{E}_t)$.
We construct a controlled sequence of such mixtures using a Norton--Bass activity process \cite{norton-bass}.
% Given a continuous synthetic progression variable $s$, and the introduction point of attack $a_i$ $\tau_i$, the attack prevalence at $s$ is defined as
Given a continuous synthetic progression variable $s$ and an attack $a_i$ introduced at $\tau_i$, the attack prevalence at $s$ is defined as
\begin{equation}
    \pi_s(a_i)
    =
    \epsilon\,\mathbf{1}[i<c(s)]
    +
    \left[1-(c(s)-1)\epsilon\right] b_i(s),
    \label{eq:attack_mixture}
\end{equation}
where $b_i(s)$ is the normalized Norton--Bass activity,
$\epsilon$ preserves probability mass for previously introduced
attacks, and
$c(s)=\max\{i:s\geq\tau_i\}$ denotes the latest introduced attack.
Each experimental environment corresponds to a selected snapshot
$s_t$, such that $\pi_t(a_i)=\pi_{s_t}(a_i)$.
The attack ordering is synthetic and is not intended to represent the
historical chronology of attack development.
\vspace{-10pt}
\section{SE-ADD}
\label{sec:method}

As shown in Fig.~\ref{fig:seadd}, SE-ADD turns the sequence defined in Eq.~\eqref{eq:model_evolution} into a self-evolution process.
At round $r$, the ALM generates forensic cues from the training samples and uses its current mistakes to build various mistake-driven supervision.
The resulting supervision is then used to update the LoRA adapter from $\phi_{t,r}$ to $\phi_{t,r+1}$.
The updated ALM regenerates forensic cues and verdicts, allowing the supervision itself to evolve with the adaptor.
After reaching the bounded evolution cycle, a development set is used to select the LoRA adaptor for the next environment.

\vspace{-5pt}
\subsection{Self-Generated Forensic Cues}
\label{sec:cues}
To expose the model's own forensic interpretation of each training sample, SE-ADD first elicits self-generated forensic cues from the current ALM. For each training audio sample $x_i \in \mathcal{D}^{\mathrm{train}}_t$,
the current ALM first generates $K$ candidate forensic cue lists using a fixed cue-generation prompt $P_{\mathrm{cue}}$:
\begin{equation}
    \mathcal{C}_i^{t,r}
    =
    \{c_{i,k}\}_{k=1}^{K},
    \qquad
    c_{i,k}
    \sim
    \mathcal{M}
    \left(
        \cdot \mid x_i, P_{\mathrm{cue}};
        \theta,\phi_{t,r}
    \right).
    \label{eq:cue_generation}
\end{equation}
Neither the ground-truth label $y_i$ nor the attack identity is
provided during generation. 
The retained set $\widetilde{\mathcal{C}}_i^{t,r}$ is subsequently used to construct mistake-driven supervision.
We refer to these outputs as \emph{forensic cues} rather than verified explanations, since $G(\cdot)$ performs text-level quality control but does not independently verify their acoustic correctness.

\vspace{-5pt}
\subsection{Mistake-Driven Supervision}
\label{sec:correction}

Rather than treating all training samples uniformly, SE-ADD adopts the current model's mistakes to build various mistake-driven supervision~\cite{an2024learningmistakesmakesllm,tong-etal-2024-llms}.
Specifically, it first locates current failures using cue-free
verdicts, and then uses the self-generated forensic cues from
Section~\ref{sec:cues} to construct additional supervision for those
mistakes. For each training sample $x_i$, the current ALM makes a cue-free
authenticity verdict
\begin{equation}
    \hat{y}_i^{t,r}
    =
    \mathcal{M}
    \left(
        x_i, P_{\mathrm{cls}};
        \theta,\phi_{t,r}
    \right),
    \label{eq:direct_prediction}
\end{equation}
where $P_{\mathrm{cls}}$ is a fixed binary classification prompt that
does not include $\widetilde{\mathcal{C}}_i^{t,r}$, and $\hat y_i^{t,r}\neq y_i$ marks a current mistake. We further define three supervision tasks $\tau$ for $x_i$:
\begin{equation}
\begin{aligned}
    \tau_i^{\mathrm{dir}}  &= (x_i\rightarrow y_i),\\
    \tau_i^{\mathrm{gen}}  &= (x_i\rightarrow
        (\widetilde{\mathcal C}_i^{t,r},y_i)),\\
    \tau_i^{\mathrm{cond}} &= ((x_i,\widetilde{\mathcal C}_i^{t,r})
        \rightarrow y_i).
\end{aligned}
\label{eq:supervision_tasks}
\end{equation}
where $\tau_i^{\mathrm{dir}}$, $\tau_i^{\mathrm{gen}}$ and $\tau_i^{\mathrm{cond}}$ supervises direct authenticity
prediction from the audio, the joint generation of forensic cues and the ground-truth verdict, and authenticity prediction conditioned on the generated forensic cues, respectively. 
Consequently, SE-ADD uses $\tau_i^{\mathrm{dir}}$ as the base supervision for every training sample, and augments current mistakes with the two cue-based tasks.
The supervision constructed for $x_i$ is
\begin{equation}
    \mathcal{T}_i^{t,r}
    =
    \{\tau_i^{\mathrm{dir}}\}
    \cup
    \begin{cases}
        \{\tau_i^{\mathrm{gen}},\tau_i^{\mathrm{cond}}\},
        &
        \hat{y}_i^{t,r} \neq y_i
        \land
        \widetilde{\mathcal{C}}_i^{t,r}\neq\varnothing,\\
        \varnothing,
        & \text{otherwise}.
    \end{cases}
    \label{eq:corrective_tasks}
\end{equation}
Since the augmentation may produce unequal numbers of training tasks across two labels, the resulting task pool for supervised fine-tuning is class-balanced by oversampling. 

\vspace{-5pt}
\subsection{Cumulative Self-Evolution}
\label{sec:evolution}

To turn mistake-driven correction into persistent self-evolution,
SE-ADD updates the current LoRA state rather than training a new
adapter at each round. With the class-balancde task pool $\mathcal{T}_{t,r}$ constructed from $\{\mathcal{T}_i^{t,r}\}_{i=1}^{N_t}$, the next LoRA parameters are updated as$
    \phi_{t,r+1}
    =
    \operatorname{SFT}
    \left(
        \phi_{t,r};
        \mathcal{T}_{t,r},
        \theta
    \right),$
with fixed $\theta$.
Hence, the adaptation is accumulated across rounds rather than reinitialized from the beginning. After $R$ rounds of the evolution within each environment, a disjoint development set $\mathcal{D}^{\mathrm{dev}}_t$ is used to select the round $r_t^{*}$ with the lowest EER:
\begin{equation}
    r_t^{*}
    =
    \arg\min_{r \in \{1,\ldots,R\}}
    \mathrm{EER}
    \left(
        \mathcal{M}(\theta,\phi_{t,r}),
        \mathcal{D}^{\mathrm{dev}}_t
    \right).
    \label{eq:round_selection}
\end{equation}
The LoRA parameters $\phi_{t,r_t^{*}}$ are then carried forward to initialize the ALM $\mathcal{M}(\theta,\phi_{t+1,0})\leftarrow\mathcal{M}(\theta,\phi_{t,r_t^*})$ afterwards.

\section{Experiments}
\subsection{Experimental Setup}
\label{sec:setup}

\noindent\textbf{Environment construction.}
\begin{figure}
    \centering
    \includegraphics[width=0.9\linewidth]{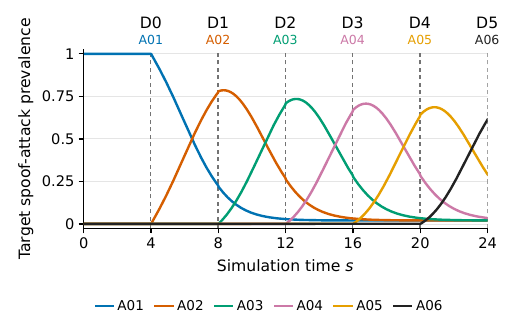}
    \caption{Constructed evolving spoofing environments.}
    \label{fig:attack_shift}
\end{figure}
We construct six environments from the ASVspoof 2019 LA train/dev partitions~\cite{asvspoof2019}, using attacks A01--A06 with the environment-specific spoof mixtures $\pi_t$ defined in Section~\ref{sec:problem}.
Each environment contains 1,024 training and 1,024 development utterances, both balanced between 512 bona fide and 512 spoof samples.
Spoof utterances are disjoint across environments, while bona fide samples are drawn from a shared split-specific pool; train and development sets are mutually disjoint. Fig.~\ref{fig:attack_shift} shows the resulting attack-prevalence trajectories and the six snapshots used to instantiate the evolving environments.
Held-out evaluation uses 8,192 balanced utterances from the evaluation partition of the ASVSpoof 2019 LA~\cite{asvspoof2019}, where spoof samples are constructed via unseen attacks A07--A19.

\noindent\textbf{Models and training.}
Two ALMs are adopted as backbones in SE-ADD: Qwen2-Audio-7B-Instruct (Qwen2-Audio)~\cite{Qwen2-Audio} and MOSS-Audio-8B-Instruct (MOSS-Audio)~\cite{mossaudio2026}.
LoRA is applied to the attention and feed-forward projections.
We use rank $8$, scaling factor $16$ for Qwen2-Audio, and dropout $0.05$,  and rank $8$, scaling factor $2$, and no dropout for MOSS-Audio.
Each environment contains two evolution rounds, and each round is trained for three epochs with a learning rate of $2\times10^{-5}$, batch size $1$, and gradient accumulation of $4$. The ALM produces $K=3$ candidate generations of forensic cue per audio, with temperature $0.8$, top-$p$ $0.95$, and maximal $160$ new tokens for each output.

\noindent\textbf{Baselines.}
Four adaptation settings are compared.
\textbf{Base} denotes the backbone ALM without adaptation.
\textbf{Direct} performs sequential supervised adaptation using only authenticity targets and follows the same environment sequence, number of rounds, training examples, and optimization steps as SE-ADD.
\textbf{Frozen-Cue} follows the SE-ADD correction procedure but reuses the self-generated forensic cues in the first round rather than regenerating them after the model update.
\textbf{SE-ADD} denotes the full method with mistake-driven supervision and cue regeneration across rounds.
Sinece cue-augmented targets are longer, \textbf{Direct} and \textbf{SE-ADD} have no identical token exposure despite their matched numbers of training examples and optimization steps.

\noindent\textbf{Evaluation Metric.}
Equal error rate (EER) is used as the primary detection metric, with lower values indicating better performance~\cite{EER}.
We report cue-free and self-hint inference where applicable to separate the detector's direct decision capability from its use of self-generated forensic cues.
\vspace{-10pt}
\subsection{Within-Environment Self-Evolution}

\begin{figure}
    \centering
    \includegraphics[width=\linewidth]{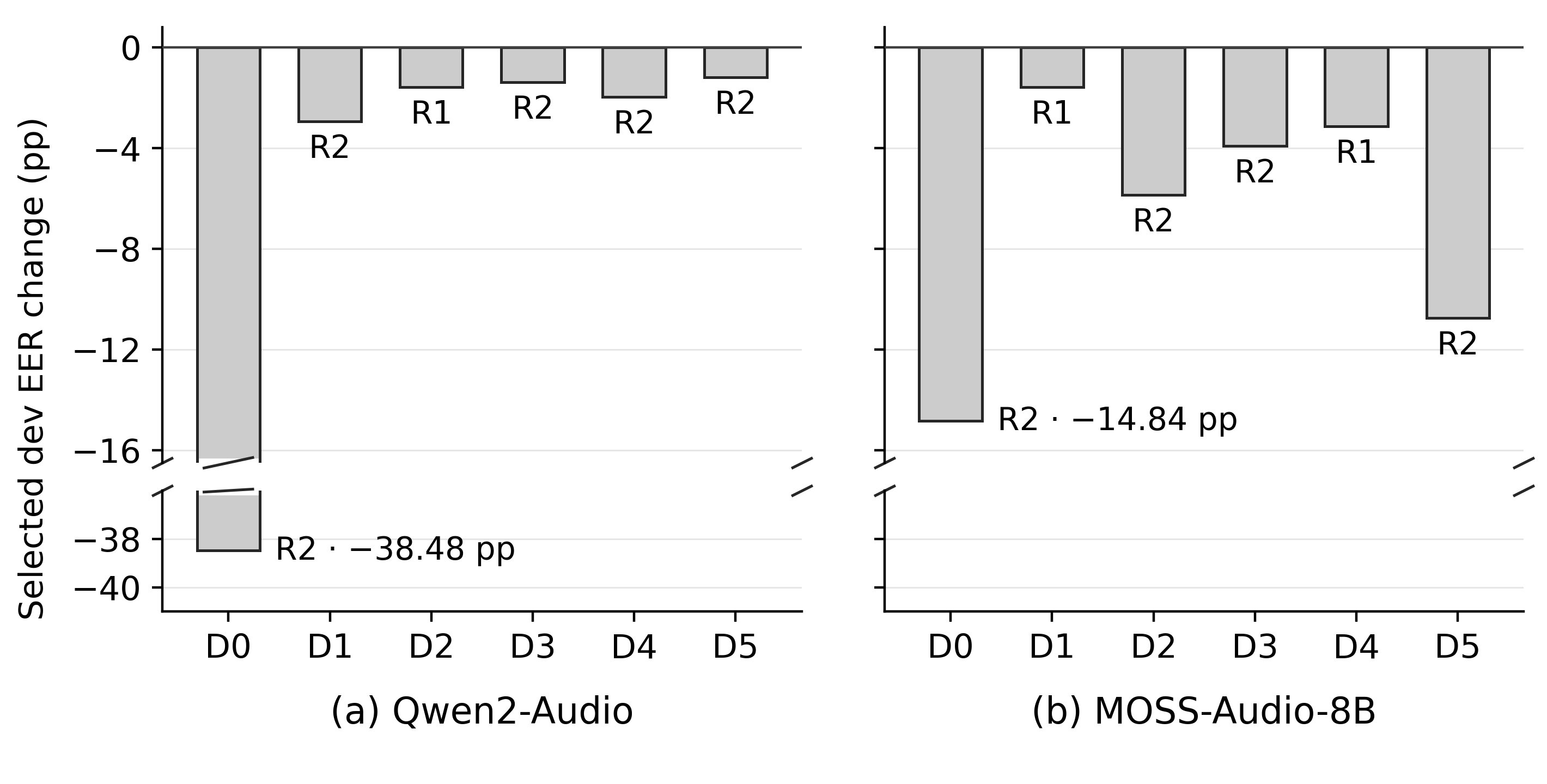}
    \caption{Within-environment development-set EER change.}
    \label{fig:within_evolution}
\end{figure}
We examine whether SE-ADD brings beneficial updates within
each environment.
Figure~\ref{fig:within_evolution} reports the development-set EER
change of the selected evolved model relative to the model entering
the same environment, where negative values indicate improvement and labels denote the selected round.
Across both ALMs, the selected model achieves a lower EER than the
incoming model in all six environments, indicating that SE-ADD identifies beneficial adaptation within each shifted spoofing mixture.
The development-selected round varies across environments: Qwen2-Audio selects R1 in $D_2$, while MOSS-Audio selects R1 in $D_1$ and $D_4$; R2 is selected otherwise.
This indicates that the preferred checkpoint is environment- and backbone-dependent.
\vspace{-10pt}
\subsection{Held-Out Generalization across Evolution}
\label{sec:generalization}
\begin{figure}
    \centering
    \includegraphics[width=\linewidth]{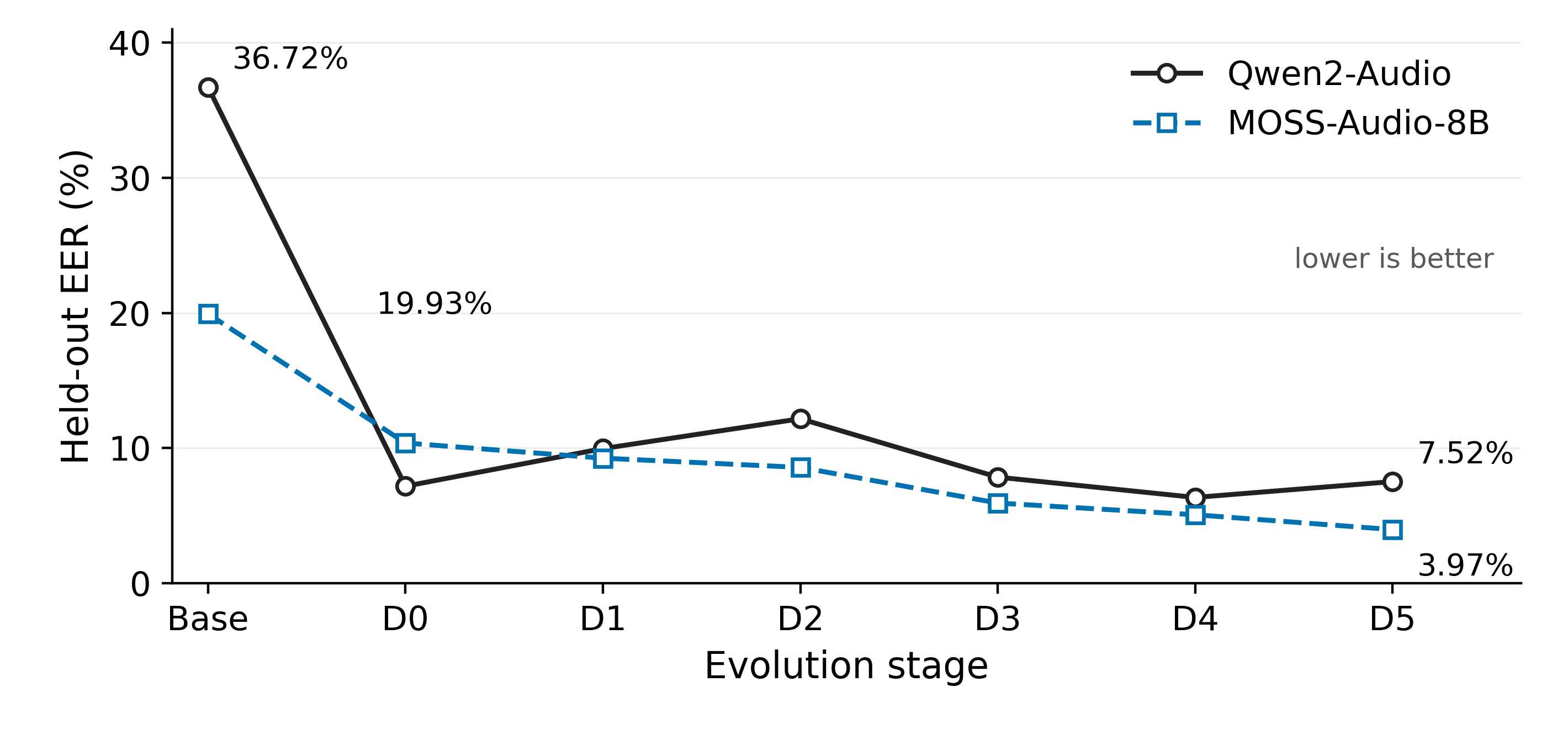}
    \caption{Held-out unseen-attack EER across evolution stages.}
    \label{fig:heldout_evolution}
\end{figure}
\vspace{-5pt}
We explore whether the adaptation accumulated across shifting spoofing environments transfers to attacks that are never observed during evolution.
As shown in Fig.~\ref{fig:heldout_evolution}, both ALMs retain substantial improvements on the fixed held-out set as evolution proceeds.
For Qwen2-Audio, the EER decreases from $36.72\%$ for the frozen base model to $7.52\%$ after the final environment, while MOSS-Audio decreases from $19.93\%$ to $3.97\%$.
The intermediate trajectories are not strictly monotonic, implying that adaptation to a newly shifted attack mixture can temporarily trade off performance on unseen attacks.
Besides, the large overall reductions from the backbones show that the knowledge accumulated over the A01--A06 stream transfers beyond the attacks used for adaptation.

We further examine retention on previously observed environments in Qwen2-Audio.
The final $D_5$ model achieves a mean EER of $2.23\%$ on $D_0$--$D_4$, although performance degrades by an average of $0.66$ percentage points relative to the best earlier checkpoints. Thus, cumulative adaptation exhibits mild forgetting while largely retaining performance on previously observed attack mixtures.

\vspace{-10pt}
\subsection{Effect of Forensic-Cue Supervision}
\label{sec:cue_effect}
\begin{table}[t]
\centering
\caption{Held-out EER (\%) under cue-free vs. self-hint.}
\label{tab:cue_effect}
\begin{tabular}{lcccc}
\toprule
& \multicolumn{2}{c}{Qwen2-Audio}
& \multicolumn{2}{c}{MOSS-Audio} \\
\cmidrule(lr){2-3}\cmidrule(lr){4-5}
Method
& Direct & Self-hint
& Direct & Self-hint \\
\midrule
Base
& 36.72 & 43.12
& 19.93 & 37.78 \\
Direct-SFT
& \textbf{6.42} & 6.45
& \textbf{3.47} & 21.07 \\
SE-ADD
& 7.52 & \textbf{5.90}
& 3.97 & \textbf{9.84} \\
\bottomrule
\end{tabular}
\end{table}
Table~\ref{tab:cue_effect} shows that the benefit of forensic-cue supervision depends on the inference mode.
Under cue-free inference, Direct-SFT achieves lower EER than SE-ADD on both backbones (6.42\% vs.\ $7.52\%$ for Qwen2-Audio and $3.47\%$ vs.\ $3.97\%$ for MOSS-Audio).
Under self-hint inference, however, SE-ADD outperforms Direct-SFT, reducing EER from $6.45\%$ to $5.90\%$ and from $21.07\%$ to $9.84\%$, respectively.
This indicates that forensic-cue supervision mainly improves the model's ability to use explicit forensic cues rather than uniformly strengthening cue-free discrimination.

However, the absolute effect of cue conditioning remains backbone-dependent. Self-hints improve SE-ADD on Qwen2-Audio from $7.52\%$ to $5.90\%$, but degrade performance on
MOSS-Audio from $3.97\%$ to $9.84\%$).
Thus, learning from forensic cues and benefiting from them at inference are related but not equivalent capabilities.

\vspace{-10pt}
\section{Conclusion}
\label{sec:conclusion}
We introduced SE-ADD, which adapts ALMs under evolving spoofing environments through mistake-driven supervision from self-generated forensic cues and current detection failures. 
Across two ALM backbones, SE-ADD yields beneficial within-environment updates and transfers to held-out unseen attacks, while the benefit of forensic-cue supervision remains inference- and backbone-dependent.

\bibliographystyle{IEEEbib}
\bibliography{strings,refs}

\end{document}